\documentstyle[epsfig, twocolumn]{iso98}

\newcommand{\thepapername}{}

\newcommand{\lesssim}{\mathrel{\hbox{\rlap{\hbox{\lower4pt\hbox{$\sim$}}}\hbox{$<$}}}}
\newcommand{\gtrsim}{\mathrel{\hbox{\rlap{\hbox{\lower4pt\hbox{$\sim$}}}\hbox{$>$}}}}

\newcommand{\et}{\it et~al.}                  %

\begin{document}

\setlength{\parindent}{0pt}
\setlength{\parskip}{10pt plus 1pt minus 1pt}
\setlength{\hoffset}{-1.5truecm}
\setlength{\textwidth}{17.1 true cm}
\setlength{\columnsep}{1truecm}
\setlength{\columnseprule}{0pt}
\setlength{\headheight}{12pt}
\setlength{\headsep}{20pt}
\pagestyle{veniceheadings}

\title{\bf ISOCAM EXTRAGALACTIC MID-INFRARED DEEP SURVEYS UNVEILING
DUST-ENSHROUDED STAR FORMATION IN THE UNIVERSE}

\author{\bf D.~Elbaz$^1$, H.~Aussel$^1$, C.J.~C\'esarsky$^1$, 
F.X.~D\'esert$^2$, D.~Fadda$^1$, A.~Franceschini$^3$,\\
\bf M.~Harwit$^4$, J.L.~Puget$^2$, J.L.~Starck$^1$ \vspace{2 mm} \\ 
$^1$Service d'Astrophysique, CEA/DSM/DAPNIA Saclay,
Orme des Merisiers, 91191 Gif-sur-Yvette C\'edex, France \\ fax
$+$33.1.69.08.65.77, e-mail {\tt elbaz@cea.fr} \\ 
$^2$Institut d'Astrophysique Spatiale, B{\^a}t 121, 
Universit\'e Paris XI, F-91405 Orsay C\'edex, France \\ 
$^3$Osservatorio Astronomico di Padova, Italy \\
$^4$511 H.Street S.W., Washington, DC 20024-2725; also Cornell University}

\maketitle

\begin{abstract}
ISOCAM extragalactic mid-infrared deep surveys have detected a
population of strong infrared emitters ten times more numerous than
expected if there were no evolution from z $<$ 0.2 (IRAS) up to the
maximum redshift of these galaxies (z$\simeq$1.5). The mid-infrared
cosmic background produced by these galaxies ($\simeq 2.35\pm
0.8~nW~m^{-2}~sr^{-1}$, at 15\,$\mu$m above 50\,$\mu$Jy) is larger
than 30 per cent of the energy radiated in the I band by the optical
galaxies detected in the HDF, which are two orders of magnitude more
numerous. This fraction is much higher than in the local universe
(z$<$0.2) as probed by IRAS, where all integrated infrared emission
from 8 to 1000\,$\mu$m makes 30 per cent of the optical starlight
(Soifer $\&$ Neugebauer 1991). Even assuming a conservative spectral
energy distribution (SED), they produce a major contribution to the
140 $\mu m$ DIRBE background measured by Hauser $\it{et~al.}$ (1998,
$\simeq 25.1\pm7~nW~m^{-2}~sr^{-1}$) and Lagache $\it{et~al.}$ (1999,
$\simeq 15.3\pm9.5~nW~m^{-2}~sr^{-1}$). This is both a confirmation of
the strong infrared cosmic background that was first detected by Puget
$\it{et~al.}$ (1996) and the first identification of the galaxies
responsible for a large fraction of this background at 140\,$\mu$m.
We were able to identify the galaxies responsible for this strong
infrared emission in the region of the Hubble Deep Field (HDF plus its
flanking fields) due to the large number of ground-based observations
of HDF galaxies. We find that these bright (several times
$10^{11}~L_{\odot}$) and massive ($<M>\simeq 1.5\times
10^{11}~M_{\odot}$) infrared galaxies have a typical redshift of
z$\simeq$0.7 and optical colors similar to field galaxies. The
fraction of galaxies with morphological signs of interactions is
larger at higher redshift. This is consistent with the nature of local
luminous infrared galaxies (LIGs, with $L_{bol}\simeq
L_{IR}>10^{11}~L_{\odot}$) found by IRAS, which often exhibit evidence
of galaxy interaction or merging (Sanders $\&$ Mirabel 1996) and
indicates that interactions should play a crucial role in the
evolution of galaxies as expected from bottom-up scenarios.
\vspace{5 pt} 
\end{abstract}

\section{INTRODUCTION}
\label{SECTION:intro}
Early-type galaxies and the central bulges of spirals are expected to
have rapidly converted most of their primordial gas into stars as
indicated by their red colors, typical of old stellar populations, and
the lack of remaining fuel to entertain new star formation. The bulk
of these stars was therefore expected to have formed during a major
episode of star formation and the distant galaxies which should have
experienced this process were called 'primeval' galaxies. The failure
of optical searches for such primeval galaxies, expected to exhibit
strong Ly-$\alpha$ emission lines (Djorgovski $\&$ Thompson 1992), was
interpreted with the help of two scenarios which are not mutually
exclusive. On one hand, the predictions of bottom-up cosmological
scenarios of galaxy formation (White $\&$ Frenk 1991) and the
observational evidence for merging galaxies, suggest that primeval
galaxies simply do not exist, because galaxy formation is a continuous
process. On the other hand, strong dust extinction could have masked
these major episodes of star formation so that primeval galaxies
should be found in infrared surveys (Franceschini $\et$ 1991, 1994).

Optical surveys found a strong evolution of the population of blue
galaxies as a function of redshift for galaxies below z=1
(Canada-France Redshift Survey, CFRS, Lilly {\it et~al.} 1995, Hammer
{\it et~al.} 1997). At larger redshifts, Steidel $\et$ (1996) used the
U and B drop out technique to select galaxies at redshifts 3 and 4
from their broad-band colors. Madau $\et$ (1996) translated these
observations into a famous plot showing the average star formation
history of the universe, which could also be seen as the universal
history of an ideal galaxy. In this plot, the formation of stars per
cubic megaparsec of the universe was expected to peak at a redshift
close to 1 and then decrease at larger redshifts. However, when
observed spectroscopically, the drop-out galaxies were found to
exhibit a flat spectrum in the UV, indicative of a strong dust
extinction which could lead to an underestimation of the associated
star formation rates (SFR) of these galaxies by a factor of three and
maybe even as much as ten (Pettini $\et$ 1997, Meurer $\et$ 1997).

In the local universe, direct infrared observations from IRAS (Soifer
$\&$ Neugebauer 1991) showed that the infrared luminosity from 8 to
1000\,$\mu$m of galaxies is about 30 per cent of that from
starlight. The population of LIGs only produce about 6 per cent of
this integrated infrared emission. Hence, although the detection by
IRAS of LIGs, which appeared to be the strongest starbursts ever
detected, was legitimately considered as a breakthrough, it was not
supposed to change our understanding of the history of star formation
because of the marginal proportion of stars born in these
galaxies. However, analyses of IRAS extragalactic source counts showed
evidence for strong evolution at low flux levels for ULIGs
(ultra-LIGs, $L_{bol}\simeq L_{IR}>10^{12}~L_{\odot}$, Hacking $\et$
1987, Lonsdale $\&$ Hacking 1989, Lonsdale $\et$ 1990). More recently,
Kim $\&$ Sanders (1998) found a very strong density evolution for
ULIGs with a 60\,$\mu$m flux between 0.5 Jy and 1.5 Jy of the form
$\Phi (z)\simeq~(1+z)^{7.6\pm3.2}$. This result was only tentative
because of the small redshift range sampled by IRAS luminous galaxies
(z$<$0.27), but it is an indication that ULIGs and maybe also LIGs
should have played a stronger role in the past. The second argument in
favor of such an evolution was found by Puget {\it et~al.}  (1996) who
detected a strong cosmic infrared background (CIRB) in the 300\,$\mu$m
to 1 mm range in the COBE-FIRAS data. This result was then confirmed
by Guiderdoni $\et$ (1997) and by a positive detection at lower
wavelength using another instrument on-board COBE, DIRBE. The exact
value of the COBE-DIRBE CIRB at 140\,$\mu$m is still a matter of
debate (Hauser {\it et~al.} 1998, find a value of
$25\pm7~nW~m^{-2}~sr^{-1}$, while Lagache {\it et~al.} 1999, find
$15\pm9~nW~m^{-2}~sr^{-1}$), but most importantly its existence is now
confirmed by independant teams and suggests that the rest-frame
infrared emission of galaxies is much higher at high redshift than
estimated locally (z$<$0.2) by IRAS.
\begin{figure}[!b]
\centerline{\psfig{file=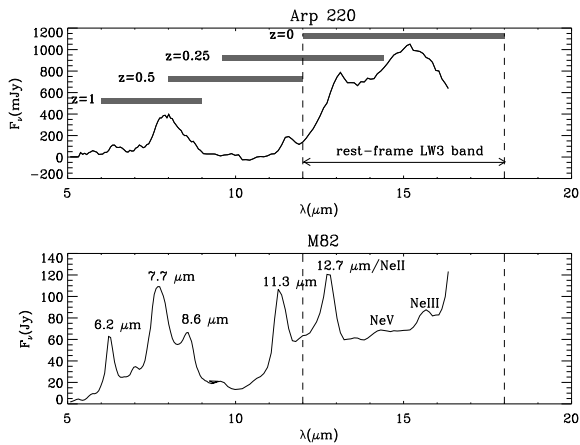,height=6.0cm,width=8.0cm}}
\caption{\em SEDs from ISOCAM Circular Variable Filter. The
displacement of the 12-18 $\mu$m (LW3) bandpass is plotted as a
function of redshift. Upper plot: Arp 220 (Charmandaris $\et$ 1998),
high LW3 over LW2 ratio. Lower plot: M82 (Tran 1998), showing strong
UIB features and low ratio of the 12-18 $\mu$m over the 5-8.5 $\mu$m
band.}
\label{FIGURE:cvf}
\end{figure}

Several cosmological surveys have been performed with ISOCAM, on-board
ISO, in directions of low zodiacal and Galactic cirrus mid-infrared
emission, ranging from large and shallow ones (several square degrees,
complete down to $\simeq$ 2 mJy) to narrow and very deep ones (a few
square arc-minutes, complete down to 50 $\mu$Jy). These surveys were
performed in the two main broad-band filters of ISOCAM: the 6.75
$\mu$m LW2 filter and the 15 $\mu$m LW3 filter, respectively centered
on the rest-frame parts of the SED which are dominated by aromatic
features (7\,$\mu$m band) and by the thermal emission of very small
dust grains (VSGs).
We will show in the following that with a thousand times better
sensitivity and sixty times better spatial resolution than IRAS,
ISOCAM mid-infrared extragalactic surveys have unveiled most of the
star formation in the universe below z$\simeq$1 and identified most of
the galaxies contributing to the COBE-FIRAS 140\,$\mu$m cosmic
background. The galaxies responsible for most of the mid-infrared
light are located at z$\simeq$0.7 and have a bolometric luminosity
larger than $10^{11}~L_{\odot}$. We will discuss this result
considering the role of interactions and comparing the relative roles
of star formation and Active Galaxy Nuclei (AGN) activities.
\section{ORIGIN OF THE MID-IR EMISSION AND K-CORRECTION}
\label{SECTION:origin}
The rest frame mid-IR emission of galaxies can be divided into three
components:

\begin{itemize}
\item
{\bf UIBs:} the Unidentified Infrared Bands (UIBs), detected at 6.2,
7.7, 8.6, 11.3 and 12.7 $\mu$m as well as their underlying continuum,
dominate the mid-IR emission below 12 $\mu$m (see
figure~\ref{FIGURE:cvf}).  The carriers of these UIBs are proposed to
be aromatic carbon species such as PAHs (Polycylic Aromatic
Hydrocarbons, L\'eger $\&$ Puget 1984, Puget $\&$ L\'eger 1989,
Allamandola $\et$ 1989) or coal grains (Papoular 1991). Below 12\,$\mu$m,
where the SED is dominated by the UIBs, the intensity of the
interstellar radiation field produces only small changes in the shape
of the spectrum. This is strong indication that the carriers of the
UIBs are transiently heated by the absorption of individual photons
(Boulanger 1998).
\item 
{\bf Warm dust (T$>$150 K):} Very Small Grains (VSGs) of dust heated
by optical-UV photons emitted by stars produce a continuum at {\bf
$\lambda > 10~\mu m$} (D\'esert $\et$ 1990).
\item
{\bf Forbidden lines of ionized gas:} NeII (12.8\,$\mu$m), NeIII (15.6
$\mu m$), SIV (10.5\,$\mu$m), ArII (7\,$\mu$m). These lines are good
indicators of the star formation activity, particularly the NeIII/NeII
ratio, which is correlated to the temperature of the stars ionizing
the ISM.
\end{itemize}

For local galaxies, the LW2 (5-8.5\,$\mu$m) and LW3 (12-18\,$\mu$m)
filters measure respectively the MIR emission due to UIBs and to VSGs
(emission lines are negligible in these broad-band filters). The LW3
band becomes more and more dominated by UIBs with increasing redshift,
due to k-correction (see figure~\ref{FIGURE:cvf}). At z$\simeq$1, the
effective LW3 band corresponds to the rest-frame LW2 band typically,
so it probes UIBs, not hot dust, and above z$\simeq$1.5, dust emission
(UIBs + VSGs) becomes too faint to be detected by ISOCAM.

\begin{figure}[]
\centerline{\psfig{file=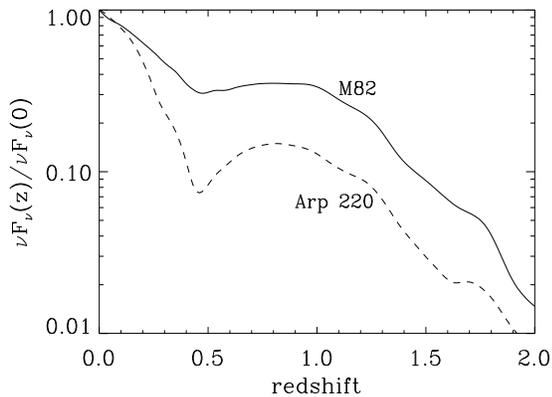,height=6.0cm,width=8.0cm}}
\caption{\em K-correction for the SED of M82 (plain line) and Arp 220
(dashed line) in the LW3 band (12-18\,$\mu$m).}
\label{FIGURE:kcor}
\end{figure}

The k-corrections for the SED of M82 and Arp 220 have been plotted in
figure~\ref{FIGURE:kcor}. A galaxy with an SED like M82, or Arp 220,
will be fainter in $\nu F_{\nu}$ with increasing redshift up to a
redshift of typically z$\simeq$0.4-0.5. Above this redshift, however,
the k-correction will become negative, i.e.\ the galaxy will appear
brighter, due to the entrance of the UIBs into the LW3 band, with a
maximum around z$\simeq$0.7. Indeed, in very deep ISOCAM surveys, like
in the HDF field, we will see that most of the galaxies are located
above z=0.4, with a mean redshift of z=0.7. Above a redshift of
z$\simeq$1.4, the k-correction falls rapidly, which explains why
there is a cut-off in the redshift distribution of ISOCAM galaxies.

The MIR emission of a galaxy is correlated with its star formation
activity as illustrated by the case of the Antennae galaxy (Vigroux
$\et$ 1996, Mirabel $\et$ 1998). Most of the star formation in this
system, as probed by ISOCAM, takes place in a source lying in the
overlapping region of the two interacting galaxies, NGC
4038/4039. This source is optically faint and most of the optical
emission arises from the two galactic nuclei. In this case, one would
underestimate the SFR when using only the optical-UV part of the
SED. It is not straightforward, however, to estimate a star formation
rate in starburst regions with good precision using only the MIR flux
of a galaxy. Indeed, the MIR results from a mixture of stochastic
heating (UIBs) and thermal emission at high temperature (VSGs), while
the FIR results from thermal emission of big grains at lower
temperature. However, assuming that the physics of local galaxies is
representative of that of more distant ones, one can still compare the
properties of local and distant galaxies in order to quantify the
evolution of galaxies.

\section{DESCRIPTION OF THE ISOCAM EXTRAGALACTIC SURVEYS}
\label{SECTION:ds}
\begin{table*}[!ht]
\begin{center}
\leavevmode
\caption{\em Description of the ISOCAM extragalactic surveys. We have
only mentionned here those surveys used to derive the log N-log S
presented in this paper.}
\vspace{0.5 em}
\begin{tabular}{llllll}
\hline \\[-5 pt]
Region & Name & area & [F$_{min}$,F$_{max}$] & t$_{exp}$ & CIRB \\
       &      &      & (mJy)              & ksec      & $nW~m^{-2}~sr^{-1}$ \\
\hline \\[-5 pt]
A2390 & Ultra-Deep Survey & 6.8 arcmin$^2$ & [0.05,0.2] & 5-25 & 1$\pm$0.4 (43 $\%$)\\
 &  &  & {\it [0.03,0.2]} & {\it 5-25} & {\it 2$\pm$0.9 (60 $\%$)}\\
HDF-North & Ultra-Deep Survey & 27.4 arcmin$^2$ & [0.1,0.3] & 6.4-19.2 & 0.7$\pm$0.3 (30 $\%$)\\
Marano Field & Ultra-Deep Survey & 49 arcmin$^2$ & [0.1,1.0] & 5.4-9.7 & 1.35$\pm$0.5 (57 $\%$)\\
Lockman Hole & Deep Survey & 484 arcmin$^2$ & [0.4,1.5] & 0.9  & 0.5$\pm$0.2 (21 $\%$)\\
Lockman Hole & Shallow Survey & 0.45 deg$^2$ & [1.2,5] & 0.18 & 0.2$\pm$0.1 (9 $\%$)\\
ELAIS fields & Very Shallow Survey & 12 deg$^2$ & [2,50] & 20 sec & 0.2$\pm$0.1 (9 $\%$)\\
\hline
\end{tabular}
\label{TABLE:parametres}
\end{center}
\end{table*}
Before the launch of ISO, there was a great deal of discussion about
the best strategy that would both optimize the instrument capabilities
and the scientific outcome of the ISOCAM mid-infrared extragalactic
surveys.  Large surveys would have to be shallow because they are time
consuming. They would probe the brightest objects at large distances
and would otherwise give good statistics for nearby objects. Pencil
beams could go much deeper, although limited by the confusion limit of
a 60-cm telescope, and could probe a comparable volume of the
universe, with a large extent over redshift or time. Because there was
no obvious answer to this question, Guaranteed Time deep surveys were
divided into a shallow survey, a deep survey and an ultra-deep survey,
and in order to be less biased by large-scale structures, it was
decided to perform these two types of surveys in the northern (Lockman
Hole) and in the southern (Marano Field) hemispheres. These fields
were obviously chosen because of their low foreground Galactic
emission (low zodiacal and cirrus emission) but also because they were
already covered at other wavelengths and in particular in the X-ray by
ROSAT and now other satellites. We will see in the next section that
this choice was beneficial.  It allowed us to achieve good statistics
over a large flux range and in particular around 1 mJy where we found
a rapid change in the slope of the number counts. The use of the
lensing magnification of galaxies in the line of sight of galaxy
clusters, like A2390 (Altieri $\et$ 1998, see also Metcalfe $\et$,
these proceedings), allowed us to complete the number counts below 100
$\mu Jy$.

These surveys were complemented by other surveys from the ISO Open
Time. The European Large Area Infrared Survey (ELAIS, european
consortium of 19 institutes lead by M.Rowan-Robinson, see these
proceedings), for sources above $\simeq$ 2 mJy, and the ISOCAM-HDF
(P.I. M.Rowan-Robinson, Rowan-Robinson $\et$ 1997, Aussel $\et$ 1999,
D\'esert $\et$ 1999) which improved the log N-log S and allowed us to
identify the galaxies responsible for the infrared excess (Aussel
$\et$ these proceedings).

Finally, the combination of all surveys is statistically significant
from 50\,$\mu$Jy up to 50 mJy and even 300 mJy, including IRAS (see
Table ~\ref{TABLE:parametres}). Hence they cover four orders of
magnitude in flux and therefore give a very strong constraint on the
evolution of galaxies in the universe below typically z$\simeq$1.4
(upper limit due to k-correction). We have only mentioned in Table
~\ref{TABLE:parametres} those surveys used to derive the log N-log S
presented in this paper. Column 3 gives the total area covered by each
survey. The depth of the surveys is not homogeneous over this total
area but the variation of the signal-to-noise ratio as a function of
the position in the mosaic was taken into account in the source
detection and in the determination of the number counts. Column 4,
[F$_{min}$,F$_{max}$], gives the flux range over which the number of
detections is statistically significant. Column 5, t$_{exp}$, gives
the exposure time per sky position, i.e.\ after co-addition of all
pixels which observed the same position on the sky. This exposure time
depends on the number of redundancies for this given position and was
summarized with two numbers when a large fraction of the image is seen
at very different depths. The last column gives the integrated
contribution of the sources detected in each survey to the 15\,$\mu$m
cosmic background, i.e.\ 2.35$\pm$0.8 $nW~m^{-2}~sr^{-1}$ above the
0.05 mJy level, or the less statistically significant value of {\it
3.3$\pm$1.3 $nW~m^{-2}~sr^{-1}$} above {\it 0.03 mJy}.  Other ISOCAM
extragalactic surveys were performed during ISO lifetime: Taniguchi
$\et$ (1997) made an ultra-deep survey of the Lockman Hole in the LW2
7\,$\mu$m band, a new ISOCAM-HDF observation was performed on the
southern HDF field (P.I. M.Rowan-Robinson, Oliver $\et$ 1999) and both
a deep and a second ultra-deep survey were also performed on the
Marano Field (P.I. C.Cesarsky). Finally, one of the CFRS fields at
1415+52 was also covered by ISOCAM at 7\,$\mu$m (Flores $\et$ 1998a)
and 15\,$\mu$m (Flores $\et$ 1998b). With a sensitivity close to the
Lockman Hole Deep Survey on a smaller area, it allowed us to identify
some ISOCAM detections (see Section~\ref{SECTION:nature}).

\section{DATA ANALYSIS}
\label{SECTION:data}
The data analysis of ISOCAM images appeared to be more complex than
expected. The main culprits for this major difficulty are the cosmic
ray impacts inducing memory effects on the detectors. It was therefore
decided to develop two different techniques which were used
independantly on the same datasets. The resulting source lists with
position and photometry were compared in order to check the robustness
of the two tools. We also decided to perform some Monte-Carlo
simulations in order to estimate the level of incompleteness and the
photometric accuracy as a function of the source fluxes. The two
techniques have been applied to the ISOCAM-HDF ultra-deep survey and
published in the same issue of A$\&$A (see Desert {\it et~al.} 1999,
for the 'three-beam technique', and Aussel {\it et al.} 1999, for the
'PRETI technique').

ISOCAM data are subject to standard gaussian noise (photon and readout
noises) and to errors associated with the flat-fielding and dark
current substraction. But the main limitation of ISOCAM deep surveys
comes from its thick and cold pixels:
\begin{itemize}
\item because they are {\bf thick}, ISOCAM pixel detectors are very
sensitive to cosmic ray impacts (4.5 pixels receive one glitch per
second). The behaviour of these glitches can be divided into three
families:
\begin{itemize}
\item "normal glitches": the more common ones, which correspond to
electrons and last only one or two readouts. They are easily removed
with a median filtering (the combination of several scales for the
median allows the best correction).
\item "faders": these glitches as well as the following ones are
probably associated with protons and alpha particles. They induce
positive peaks in the detector response, which can last several
readouts. Since ISOCAM is best used in the raster mode, a real
source will look like these glitches, i.e.\ a positive response over
the number of readouts spent on a given position of the sky.
\item "dippers": some glitches are followed by a trough extended
over more than one hundred readouts.
\end{itemize}
\item ISOCAM pixels are {\bf cold}, so that electrons move very slowly
within them and therefore induce a transient behaviour: a pixel will
take several hundred readouts to stabilize when moving from the
background to the position of a source on the sky and inversely.
Because of time limitation, one is therefore limited to non stabilized
signals which results in an uncertainty on the photometry.  This
uncertainty is strongly reduced by the partial correction of the
transient behaviour and by the use of simulations to define a
statistical distribution of measured fluxes for any given input flux.
The final uncertainty on ISOCAM deep survey fluxes is on the order of
20 per cent.
\end{itemize}

In order to facilitate the separation of sources from cosmic
ray impacts, ISOCAM surveys were performed using the raster
mode with a redundancy (number of different pixels falling
successively on a given sky position) ranging from 2 for the
shallowest survey (ELAIS) to 88 for the deepest surveys (Marano
Field Ultra-Deep survey; 64 for the HDF field).

\subsection{Data reduction techniques}
\label{SECTION:techniques}
Four different techniques applied to the ISOCAM-HDF (North) image have
been compared: the 'Saclay technique', called PRETI (Starck $\et$
1998), the 'Orsay technique', called three-beam technique (see
D\'esert $\et$ 1999), the 'Imperial College technique' (see Serjeant
$\et$ 1997) and the 'Carlo Lari technique' (Lari 1999).  After
comparison, all four techniques reach the same result at 15\,$\mu$m,
and find a small difference at 7\,$\mu$m. This will be discussed in a
common paper that will be submitted soon. The use of simulated ISOCAM
images was of great help in this work.

In the following, we give a brief description of the tool that we have
developed at Saclay, PRETI. This tool is based on a multi-resolution
wavelet transform which separates the temporal history of each
individual pixel into several timelines, each associated with a given
frequency in the variation of the signal. This technique developed by
J.L.Starck consists in searching for patterns in the wavelet space (or
frequency space), which correspond to the 'bad' glitches, namely the
faders and the dippers.  Indeed, being extended, a bad glitch will be
detected over several successive scales and appear as a pattern in the
wavelet, or frequency, space. After having isolated this pattern, one
can then substract it from the original signal, which in fact means
subtracting a smooth function without the high-frequency contribution
corresponding to the gaussian noise of the detector and where faint
sources are hidden. When the data cubes have been cleaned of their
glitches, they can be co-added to produce a mosaic image on which
standard source detection techniques can be applied. This final step
was done using again a multi-resolution wavelet transform, applied
spatially instead of temporally. The techniques (1) and (2) were
applied to all surveys: they agree on the photometry at the 20 per
cent level and the differences in astrometry are less than the pixel
size.

\subsection{Simulations}
\label{SECTION:simulations}
We performed simulations in order to quantify the sensitivity limit
(minimum detected flux, below the completeness limit), the
completeness limit (flux above which all sources are detected) and the
photometric accuracy. These simulations were performed with real
datasets (in order to simulate realistic glitches) in which we
introduced fake sources including the PSF and their modeled transient.
For this, we used a long staring observation of more than 500 readouts
and analyzed it as if it was a mosaic. In a raster observation, a real
source will only be detected when the camera points toward its
direction, while the sources present in a staring observation will
remain present over the whole observation and hence will be removed as
part of the low frequency component of the signal. We added to this
simulated raster containing real ISOCAM noise some simulated sources
with their PSF and transient behavior and then analyzed the data with
the different techniques described above. The simulations were used to
set the parameters of the data reduction technique and then to
estimate the rate of detections of sources per flux bin as well as the
error bar on the photometry. We used them to calculate the 90 per cent
confidence level error bars shown in the log N-log S
(figure~\ref{FIGURE:logN}).
\section{NATURE OF THE GALAXIES DETECTED IN ISOCAM EXTRAGALACTIC SURVEYS}
\label{SECTION:nature}
The detailed study of the properties of ISOCAM galaxies is only
beginning but with the help of the large number of multi-wavelength
observations of the HDF (Williams $\et$ 1996) and of the CFRS field at
1415+52 (Lilly $\et$ 1995), the second most observed field at all
wavelengths after the HDF, one can already begin this work. 

In the CFRS field, Flores $\et$ (1998b) have found that two thirds of
the galaxies detected by ISOCAM at 15\,$\mu$m down to a flux density
of 250\,$\mu$Jy were starburst galaxies as probed by their radio
emission at 6 and 21 cm. They recalculated the history of the metal
production or star formation per unit volume of the universe, plotted
by Madau (1996) and Lilly (1995) using only the optical and UV
information on CFRS galaxies, for the dust extinction and found a
correction factor of about three. Hence, optical-UV light would only
probe 25 per cent of the star formation in the universe below
z=1. This strong uncertainty on this correction factor needs to be
reduced by more statistics not only on infrared galaxies but also on
template SEDs of local galaxies, since the database of Schmitt $\et$
(1998) that they used only contains 59 galaxies. However, the excess
that we find in the number counts (see Section~\ref{SECTION:counts})
and the strong 15\,$\mu$m cosmic background (see
Section~\ref{SECTION:cosmic}), confirm that this plot should indeed be
corrected by a non negligible factor for dust extinction.

In the ISOCAM-HDF, we identified 36 sources, among
the 44 detections at 15 $\mu m$, with an optical counterpart from the
catalog of Barger $\et$ (1999) limited to I$<$22.5. The mean redshift
of these galaxies is $<z>\simeq 0.7$ and most of them are located
between z=0.4 and 1.4 (as expected from the k-correction, see
Section~\ref{SECTION:origin}). The mean ratio of their effective
fluxes at 15\,$\mu$m over the K band is close to that found for M82,
if it was redshifted at z=0.7.  Hence, we used the SED of M82 to
determine the rest-frame 15\,$\mu$m flux of these galaxies, although
we are aware of the strong uncertainty associated with this
estimate. The mean 15\,$\mu$m luminosity of ISOCAM-HDF galaxies, is
about ten times larger than that of M82 so this assumption should be
quite conservative since ISOCAM-HDF galaxies should be even more
active than M82 and closer to LIGs and ULIGs. Assuming that ISOCAM-HDF
galaxies present SEDs similar to that of M82, their mean infrared
luminosity would be of the order of $3\times10^{11}~L_{\odot}$ and
they would be classified as LIGs (Sanders $\&$ Mirabel 1996). For a
typical ratio of M/L$_{K}$ of about 1.5 (Charlot, private
communication), they would have a mean mass of
$1.5\times10^{11}~M_{\odot}$.

\subsection{Origin of the infrared emission: star formation versus AGN}
\label{SECTION:origin}
The major source for the huge energy radiated by these galaxies should
be star formation activity. Indeed, only a few per cent of AGNs have
been identified optically, in the radio or in the X-ray in our sample
of ISOCAM galaxies (see also Aussel $\et$, these proceedings). Genzel
$\et$ (1998) found that only 20-30 per cent of the energy radiated by
local ULIGs is powered by AGN activity, and this fraction should be
even lower for our galaxy sample since they are mainly LIGs. However,
the fraction of galaxies harboring an AGN (but not dominated by its
radiation) should be much larger: Genzel $\et$ (1998) detected the
presence of an AGN in at least 50 per cent of the ULIGs that they
studied. It was recently suggested by Fabian $\&$ Iwasawa (1999) that
if the hard X-ray background detected around 30 keV by HEAO1 were
produced by dusty AGNs (Seyferts 2) as indicated by its flat slope,
then these dusty AGNs should produce a contribution to the CIRB of
$\simeq 3~nW~m^{-2}~sr^{-1}$, i.e.\ between 10 and 20 per cent of the
CIRB measured by COBE-DIRBE. Such a contribution of AGNs to the CIRB
would be reasonable if most of it was produced by LIGs and ULIGs and
ISOCAM surveys seem to favor this option.
%
\subsection{Optical properties of the faint 15\,$\mu$m galaxies}
\label{SECTION:properties}
Using available data on the HDF, especially in the optical and NIR
from Barger $\et$ (1999), we find that the optical colors of the
galaxies detected above 100 $\mu$Jy (ISOCAM-HDF completeness limit,
Aussel $\et$ 1998) do not strongly differ from field galaxies (see
Aussel $\et$, these proceedings). Hence it would not have been
possible to discriminate the ISOCAM galaxies from their optical colors
only.

The study of the spectroscopic properties of these galaxies is still
underway (spectra made available to the community by the group of the
University of Hawaii, Barger $\et$ 1999). However, we have already
found that a large fraction of ISOCAM-HDF galaxies exhibit weak
emission lines and strong Balmer H$_\delta$ absorption lines
characteristic of the presence of a large number of A stars with an
age of typically one Gyr. On the other hand, galaxies with strong
emission lines detected in the field are not detected at 15 $\mu
m$. We also found the same result in the list of spectra obtained
after a multi-spectroscopic follow-up of 15\,$\mu$m ISOCAM detections
in the field of a nearby galaxy cluster, A1689 (P.I. P.A.Duc, see
Fadda $\et$, these proceedings). Flores $\et$ (1998b) had already
found the same result on the CFRS field at 1415+52 and interpreted
this result as a proof that the galaxies with dust-enshrouded star
formation detected by ISOCAM have been forming stars during the last
few 10$^8$ years. The lack of detection by ISOCAM of emission line
galaxies was suggested to be due to the low metallicity of these
galaxies. A completely independant spectroscopic survey of
post-starburst galaxies in clusters of galaxies located at redshifts
z=0.4-0.5 seems to already confirm this result (Dressler $\et$ 1999,
Poggianti $\et$ 1999). They propose that galaxies with weak or no
emission lines but strong Balmer $H_{\delta}$ absorption lines, that
they call e(a), are most probably dust-enshrouded starbursts using
only optical arguments based, for example, on the ratio of the OII
over H$_\alpha$ equivalent widths.
\begin{figure}[!ht]
\centerline{\psfig{file=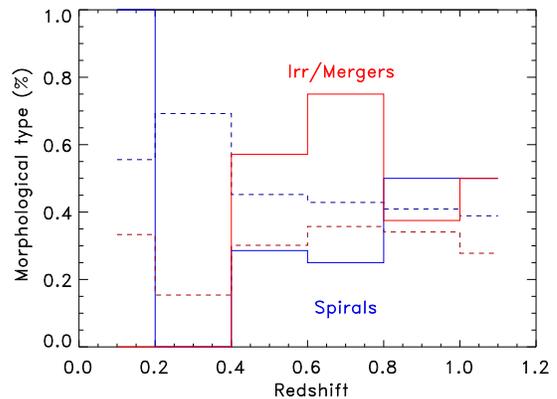,width=8.0cm,bbllx=1,bblly=1,bburx=685,bbury=505}}
\caption{\em Fraction of Peculiar versus Spiral galaxies in the
HDF$+$Flanking fields for galaxies detected at 15 $\mu$m (plain lines)
and all galaxies (dashed lines). Above z$\simeq$0.4, peculiar galaxies
dominate the distribution of galaxies detected with ISOCAM while
optically selected galaxies are mainly spirals. This population of
peculiar galaxies may produce the rapid evolution below the 1 mJy
level found in the 15 $\mu$m number counts.}
\label{FIGURE:morpho}
\end{figure}
A possible scenario for the origin of these dusty starbursts galaxies,
found both in ISOCAM extragalactic surveys and in the optical
spectroscopic surveys of distant clusters, could be linked to the
effects of galaxy interactions. Indeed, Poggianti $\&$ Wu (1999) find
an exceptionally high fraction of e(a) spectra among local LIGs where
the fraction of interacting galaxies is very high. ISOCAM detections
of CFRS galaxies also present morphological signatures of interactions
and we find a morphological segregation with increasing redshift in
the ISOCAM-HDF with more interacting galaxies with increasing redshift
(see figure~\ref{FIGURE:morpho}).
%
\section{NUMBER COUNTS}
\label{SECTION:counts}
\begin{figure}[!ht]
\centerline{\psfig{file=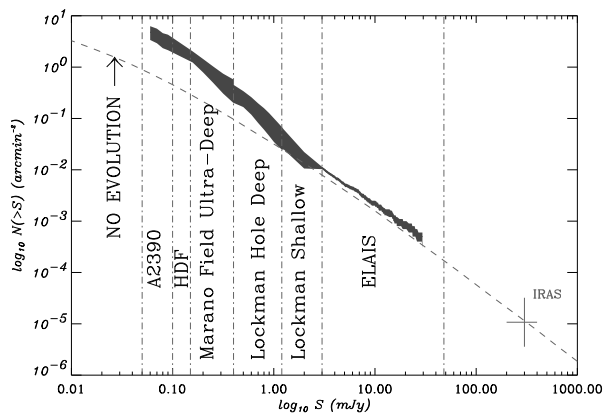,width=8.0cm,bbllx=54,bblly=362,bburx=556,bbury=720}}
\caption{\em log N($\>$S)- Log S: number of objects, N, detected at
15\,$\mu$m above a given flux level, S(mJy). The data are consistent
although they come from 6 different origins (see Table 1). Dashed
line: no evolution from IRAS (Franceschini $\et$ 1998).}
\label{FIGURE:logN}
\end{figure}

The most impressive result after three years of data analysis and
calibration is the consistency of all different surveys as shown by
the log N-log S plot over four orders of magnitude in flux density
when including the IRAS point (figure~\ref{FIGURE:logN}). The number
counts are perfectly fitted by a no-evolution model normalized to IRAS
(Franceschini 1998) above the 1 mJy level typically, while they
strongly diverge from this law below 1 mJy, with an increasing
difference which reaches a factor of 10 at the faintest fluxes (around
50\,$\mu$Jy). Although k-correction can be negative as shown in
figure~\ref{FIGURE:kcor} and therefore could explain a part of the
excess at low fluxes, it can certainly not reproduce an excess of a
factor of 10 as we find here. The detailed modeling of this evolution
is quite complex and a paper will be devoted to this subject
(Franceschini $\et$, 1999). However, the increasing fraction of
interacting galaxies as a function of redshift detected in the
ISOCAM-HDF (figure~\ref{FIGURE:morpho}) may explain the behavior of
the log N-log S, if LIGs were to begin to dominate the number counts
above z$\simeq$0.4 and completely dominate them at z$\simeq$0.7. There
was already a hint of such a strong evolution of LIGs and ULIGs in the
analysis of local ULIGs by Kim $\&$ Sanders (1998, see
Section~\ref{SECTION:intro} Introduction).

We learned from the previous section about the nature of the galaxies
that were detected in ISOCAM surveys that this excess is produced by a
few LIGs, two orders of magnitude less numerous than optical galaxies
detected in the HDF (see Aussel $\et$ 1998). Hence, the luminosity
function of galaxies should be different at z$\simeq$0.7, the mean
redshift of ISOCAM-HDF galaxies, than locally, favoring luminous
galaxies. The flattening of the log N-log S at low fluxes implies that
we are not far from having identified most of the galaxies producing
the mid-infrared background and indicates that we should be able to
give a good constraint on the cosmic background produced by these
sources.

\section{COSMIC INFRARED BACKGROUND}
\label{SECTION:cosmic}
Integrating the 15 $\mu$m number counts over the whole flux range, one
finds a conservative value of about $2.35\pm0.8~nW~m^{-2}~sr^{-1}$
above 50\,$\mu$Jy and 3.3$\pm$1.3 $nW~m^{-2}~sr^{-1}$ above
30\,$\mu$Jy, with less confidence.  This corresponds to 30-45 per cent
of the cosmic background seen with the HST in the I-band
($7~nW~m^{-2}~sr^{-1}$, Lagache $\et$ 1999). This fraction is very
high since locally IRAS showed that it is the total infrared
luminosity of galaxies, i.e.\ from 8 to 1000\,$\mu$m, which is about
30 per cent of that from starlight (Soifer $\&$ Neugebauer
1991). Assuming a conservative SED one would then expect these
galaxies to emit at least as much energy in the far infrared than in
the optical. In particular, one can estimate the effective ratio of
their 140\,$\mu$m over 15\,$\mu$m energy densities at the mean
redshift of 0.7 and compare it to the COBE-DIRBE value. A conservative
ratio would be of the order of three, like for M82, although a more
realistic one would be higher than that. In this conservative case,
they would produce more than 50 per cent of the COBE-DIRBE CIRB found
by Lagache $\et$ (1999, $\simeq 15.3\pm9.5~nW~m^{-2}~sr^{-1}$) and
about 30 per cent of the value found by Hauser $\et$ (1998, $\simeq
25.1\pm7~nW~m^{-2}~sr^{-1}$). This is only a lower limit since a ratio
of 3 is probably too conservative and since the number counts have not
completely flattened yet. The minimum ratio of the energy densities at
140\,$\mu$m over 15\,$\mu$m is about 10, hence if the faint 15\,$\mu$m
galaxies were moderate ULIGs, they would produce the whole COBE-DIRBE
background.
\section{CONCLUSION}
\label{SECTION:conclusion}
ISOCAM extragalactic surveys are nearly four orders of magnitude more
sensitive than IRAS in the 15\,$\mu$m range. They teach us that a few
luminous and massive infared galaxies located at a mean redshift of
0.7 produce most of the 15\,$\mu$m and probably also DIRBE-140\,$\mu$m
cosmic background. Assuming a conservative spectral energy
distribution for these galaxies, one find that they emit as much
energy in the infrared than the two orders of magnitude more numerous
optical galaxies detected in the HDF. The galaxies producing this
infrared excess present morphological signs of interaction. Hence,
both dust extinction and galaxy interactions played a major role in
galaxy evolution since z$\simeq$1. 

Additional conclusions will be reached only after having collected a
statistically larger sample of optical counterparts to ISOCAM
galaxies, in particular, in the ultra-deep surveys performed in the
Marano Field where we have detected hundreds of very faint 15 $\mu m$
galaxies. The large amount of energy released by these galaxies will
be one of the key questions for the next spatial telescopes, like
SIRTF and FIRST, and at larger redshifts for the ground-based
sub-millimeter interferometer LSA-MMA. The Next Generation Space
Telescope (NGST) would give a unique insight into the origin of their
activity, and in particular the proportion of the energy due to AGN
activity, if it were to observe up to 25-30\,$\mu$m.

\section*{ACKNOWLEDGMENTS}
We would like to thank Suzanne Madden for very fruitful scientific
discussions and for her advises in the redaction of this paper.


\end{document}